\documentclass[reprint,amsmath,amssymb,aps,prstab]{revtex4-1}
\usepackage[english] {babel}
\usepackage{amsfonts,hyperref,graphicx}
\usepackage{epstopdf}
\usepackage{xcolor}

\DeclareMathOperator*{\argmin}{arg\,min}

\hypersetup{
  pdfauthor={S.S. Baturin}
}

\begin{document}
\title{Stability condition for the drive bunch in a collinear wakefield accelerator}

\author{S.S. Baturin}%
\email{s.s.baturin@gmail.com}%
\affiliation{The University of Chicago, PSD Enrico Fermi Institute, 5640 S Ellis Ave, Chicago, IL 60637, USA}%
\author{A. Zholents}%
\email{azholents@anl.gov}
\affiliation{Argonne National Laboratory, Lemont, IL, 60439, USA}%
\date{\today}

\begin{abstract}
The beam breakup instability of the drive bunch in the structure-based collinear wakefield accelerator is considered and  a stabilizing method is proposed. The method includes using the specially designed beam focusing channel, applying the energy chirp along the electron bunch, and keeping energy chirp constant during the drive bunch deceleration. A stability condition is derived that defines the limit on the accelerating field for the witness bunch. 
\end{abstract}

\maketitle

\section{Introduction}\label{sec:intr}
In the collinear wake field accelerator (CWA) proposed by Voss and Weinland  in 1982 \cite{GVoss} the drive electron bunch generates the electromagnetic field by interacting with the retarding medium, typically formed either by the dielectric lined waveguide or the waveguide with small corrugations. This field, known as the wakefield, accelerates electrons of the witness bunch located at a strategically chosen distance behind the drive bunch with the maximum accelerating field and decelerates electrons of the drive bunch. The charge of a witness bunch is much smaller than the charge of a drive bunch. This promising method of particle acceleration attracted many followers (e.g., see \cite{RBriggs}-\cite{Baturin} and references therein) who pursued accelerator designs for a Linear Collider (e.g., see \cite{Jing} and reference therein) and a free-electron-laser-based light source \cite{Zhol}. A comprehensive review of the entire field of structure-based wakefield accelerators  was recently published in the Reviews of the Accelerator Science and Technology \cite{Jing1}.  Among various challenges associated with practical designs of the high energy gain and high energy efficient CWA, the one that stands out because of its extreme difficulty and importance is the task of restraining the beam breakup instability (BBU) caused by the transverse wakefields. This instability mostly affects the high charge drive bunch. The witness bunch is much less amenable to this instability because of a smaller charge and a higher energy.

Initially studied in a set of seminal papers (\cite{Panofsky}-\cite{CRY}), the BBU has been a subject of many more investigations (e.g., see \cite{BNS}-\cite{ Delayen3}). An elegant method to control the instability was proposed by Balakin, Novokhatski, and Smirnov in \cite{BNS} and was named the BNS damping thereafter. They proposed a systematic cancelation of the defocusing force of the transverse wakefield by the chromatic dependence of the quadrupole magnet strength on energy. Indeed, a key provision of the BNS damping is creation of the linear energy variation (chirp) along the electron bunch with the head electrons having higher energy. By adjusting the magnitude of the chirp, one can obtain a condition when collective betatron oscillations of electrons in each longitudinal slice of the electron bunch almost exactly repeat oscillations of the preceding slices including the head slice. In this case the frequency shift of the betatron oscillations, due to chromaticity of external focusing provided by the guiding FODO channel, balances the impact of the transverse wakefield. Here F stands for the focusing lens, D for the defocusing lens, and O for the drift space. Ultimately, the BNS damping not only guarantees the stability of the electron bunch motion, but also the preservation of the electron bunch's projected emittance. More information about the BNS damping can be found in (\cite{Balakin}-\cite{Delayen3}) and in the Chapter 3 of the textbook \cite{Chao}.

 The BNS damping gives excellent results when the wakefield is relatively small. However, in the case of the large wakefields that are typical for CWA, the BNS damping gives  an incorrect prescription for the energy chirp and fails to stabilize the BBU. A new formalism applicable for an arbitrary wakefield is proposed in this paper. 
 
We begin the analysis by considering the exact equation that describes the motion of one drive bunch electron in the structure based CWA that embedded in the FD channel. The electron is decelerated by the longitudinal wakefield and kicked by the transverse wakefield induced by other electrons of the drive bunch ahead of it. In the next step we define a special condition for electron focusing in the FD channel and transform the equation of motion to the inhomogeneous Hill's equation without the dissipation term using new variables. This leads to the definition of the first set of stability conditions for the drive bunch that includes two provisions, i.e., a requirement for the energy chirp and a requirement for preservation of the relative magnitude of the energy chirp independent of the deceleration of the drive bunch. We note that the Hill's equation is principally different from the approximate  harmonic oscillator equation considered in \cite{BNS} and other papers cited above because of the islands of instability in the particle motion it describes. Using a two particle model, we solve the inhomogeneous Hill's equation for the second particle in one FD cell and obtain bounding conditions on energy chirp corresponding to its stable motion. After that we generalize the solution to the entire electron bunch and show how the stability condition connects together key parameters of the CWA, i.e., the maximum attainable accelerating field for the witness bunch, maximum attainable gradient of the magnetic field for the lens of the FD channel, and maximum attainable energy chirp for the drive bunch. 

\section{Adaptive energy chirp and adaptive beam focusing}\label{sec:2}
\begin{figure}[h]
\begin{center}
\includegraphics[scale=0.2]{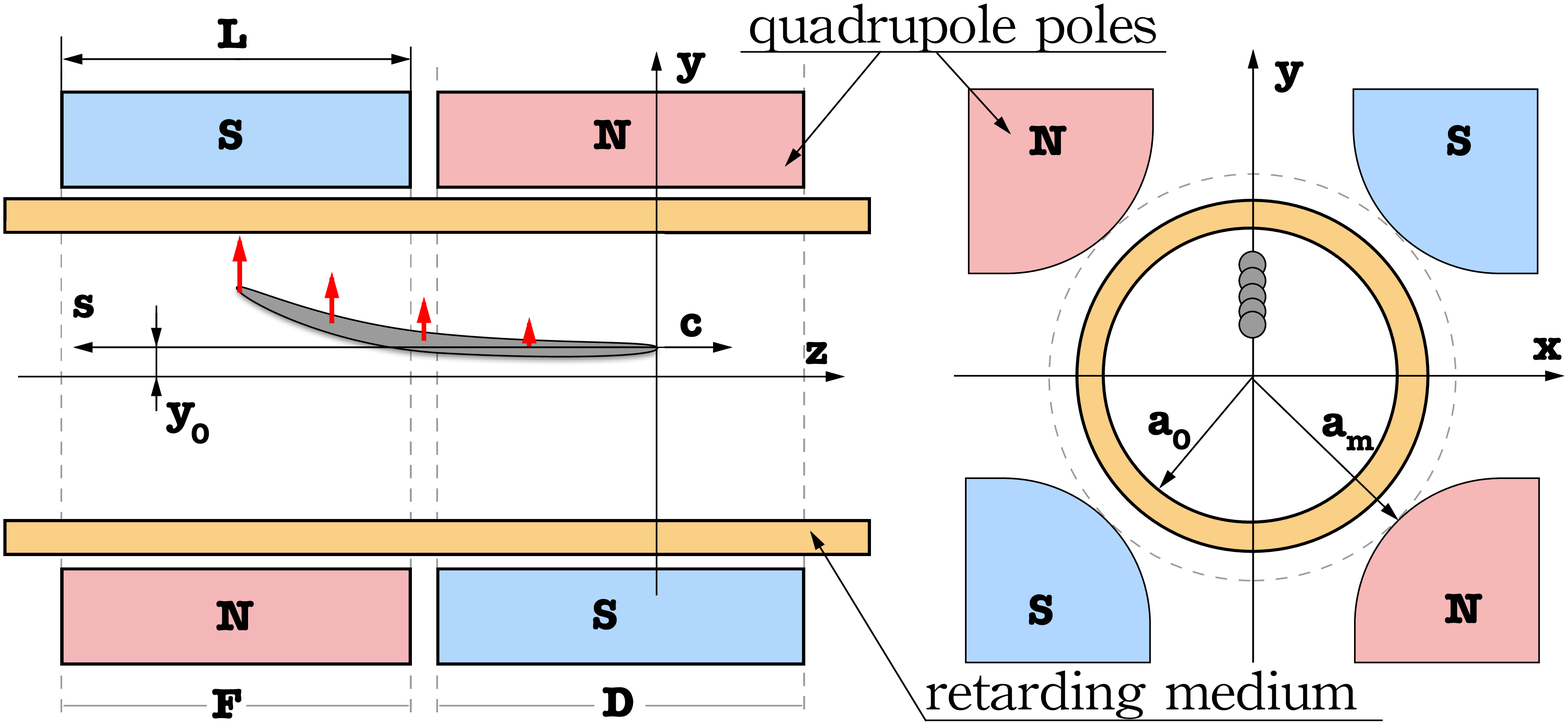}
\caption{A schematic diagram of the CWA embedded in the FD channel and the electron bunch propagating from the left to right.}
\label{Fig:1}
\end{center}
\end{figure}

We consider the relativistic drive bunch with the transverse dimensions that are much smaller than the radius $a_0$ of the wake inducing structure of the CWA. Therefore, we assume that the transverse wakefield is uniform across the bunch. In this case, the analysis of the BBU is reduced to the analysis of the stability of motion of the centers of the bunch slices located at each coordinate $s$ measured from the bunch head when the drive bunch propagates along the longitudinal axis $z$ of the CWA. We study only vertical displacements $y(s,z)$ because the analysis of the horizontal displacements $x(s,z)$ is exactly the same. 
It can be shown that the best option for a focusing system is the sequence of focusing and defocusing lenses without drift spaces that resembles a quadrupole wiggler.
We consider CWA embedded into the FD channel (see Fig.\ref{Fig:1}) with the magnetic field gradient $g(z)=\pm g_0$, where the plus sign is for the focusing (F)  lens and the minus sign is for the defocusing (D) lens. 

The evolution of the $y(s,z)$ along the CWA is described by the following equation:
\begin{align}
	\label{eq:MainDL}
	&\frac{d}{dz}\left(\gamma(s,z)\frac{dy(s,z)}{dz}\right)+\frac{e}{m_e c} g(z)y(s,z)= \nonumber\\ &=\frac{e}{m_e c^2} \int\limits_0^s G_1 (s-s_0)q(s_0)y(s_0,z) ds_0,
\end{align}
where $e$ is the electron charge, $\gamma$ is the electron energy in units $m_e c^2$, $m_e$ is the electron mass at rest, $c$ is the speed of light, and $G_1$ is the transverse Green's function. 

We define charge distribution in the electron bunch as $q(s)$ and consider bunches localized on the interval $0\leq{s}\leq{l}$ such as
\begin{align}
	\label{eq:dens}
	\int\limits_0^lq(s)ds=Q,
\end{align}
where $Q$ - is the total bunch charge.
We also assume that there is no longitudinal focusing and ignore small changes in the longitudinal velocity on the length of the CWA. This is typical for the relativistic energies. 
Under assumptions above we write:
\begin{align}
\label{eq:gmeq}
\frac{d \gamma (s,z)}{dz}=\frac{e}{m_e c^2} E_z(s,z).
\end{align}
Here $E_z(s,z)$ is the decelerating wakefield. We further define the variation of the electron energies along the bunch as 
\begin{align}
\label{eq:gm0}
\gamma(s,0)=\gamma_0 \left[1-f(s)\right]
\end{align}
and consider the bunch with the charge distribution that produces the wakefield inside the bunch equal to
\begin{align}
\label{eq:Ez}
E_z(s,z)=E_0 \left[1-f(s)\right].
\end{align}
Here $\gamma_0$ is the energy of electrons at the head of the bunch and at the beginning of the CWA and $E_0$ is the wakefield at the head of the bunch.
From \eqref{eq:gmeq},  \eqref{eq:gm0} and \eqref{eq:Ez} we get
\begin{align}
\label{eq:gmZ}
\gamma(s,z)=\gamma_0\left[1-\alpha z\right]\left[1-f(s)\right],  
\end{align}
where we define
\begin{align}
\alpha=\frac{|e|E_0}{\gamma_0m_ec^2}.
\end{align}
Thus the function $f(s)$ is equal
\begin{align}
\label{eq:dGG}
f(s)=-\frac{\gamma(s,z)-\gamma_z}{\gamma_z}=-\frac{\Delta \gamma(s,z)}{\gamma_z},
\end{align} 
where we use $\gamma_z \equiv \gamma(0,z)=\gamma_0\left[1-\alpha z\right]$ for the energy of the head of the bunch.

Satisfying Eq.\eqref{eq:dGG} requires the implementation of the adaptive energy chirp when the relative magnitude of the chirp remains constant with the deceleration of the drive bunch, while its absolute value significantly decreases. This can be done by employing the longitudinal wakefield produced by the drive bunch with a special charge distribution. For example, the ``door step" electron density distribution proposed in \cite{KBane} and considered in \cite{Baturin} gives a quasi-uniform decelerating wakefiled inside the drive bunch. Adding a small quadratic component to a linear ramp in the peak current will add a small linear variation to the decelerating field, enough to keep constant the relative magnitude of the chirp.

We notice that if we substitute $\gamma(s,z)$ in \eqref{eq:MainDL} as given by \eqref{eq:gmZ} and make a change of the variable $y(s,z)=\tilde{v}(s,\tilde{u})/\sqrt{\tilde{u}}$, where $\tilde{u}=\sqrt{1-\alpha z}$,
we arrive at
\begin{align}
	\label{eq:hillf}
	&\frac{\alpha^2 [1-f(s)]}{4}\left(\tilde{v}''(s,\tilde{u})+\frac{\tilde{v}(s,\tilde{u})}{4 \tilde{u}^2}\right) +\frac{e g(z(\tilde{u}))}{\gamma_0 m_e c} \tilde{v}(s,\tilde{u}) \nonumber\\ 
	&=\frac{e}{\gamma_0m_e c^2} \int\limits_0^s G_1 (s-s_0)q(s_0)\tilde{v}(s_0,\tilde{u}) ds_0.
\end{align}

In the region where $\tilde{u}^2>0.1$, i.e., where the drive bunch has more than $10\%$ of the initial energy, Eq.\eqref{eq:hillf} can be reduced to 
\begin{align}
\label{eq:hill_r}
&\tilde{v}''(s,\tilde{u})+\frac{4e}{\alpha^2\gamma_0 m_e c} \frac{g(z(\tilde{u}))}{1-f(s)} \tilde{v}(s,\tilde{u})= \nonumber\\ &=\frac{4e}{\alpha^2\gamma_0m_e c^2} \frac{\int\limits_0^s G_1 (s-s_0)q(s_0)\tilde{v}(s_0,\tilde{u}) ds_0}{1-f(s)}
\end{align}
assuming that
$\frac{4|e| g_0}{\alpha^2 \gamma_0 m_e c (1-f(s))}>>\frac{1}{4\tilde{u}^2}$. 

We notice that if $g(z(\tilde{u}))$ is a periodic function of $\tilde{u}$, then Eq.\eqref{eq:hill_r} is the Hill's equation.
Periodicity of $g(\tilde{u})$ can be obtained by adjusting the length of the lenses according to
\begin{align}
\label{eq:Ll}
L=L_0 \sqrt{1-\alpha z}, 
\end{align}
where $L_0$ is the length of the lens at the beginning of the CWA.
\begin{figure}[h]
\includegraphics[scale=0.2]{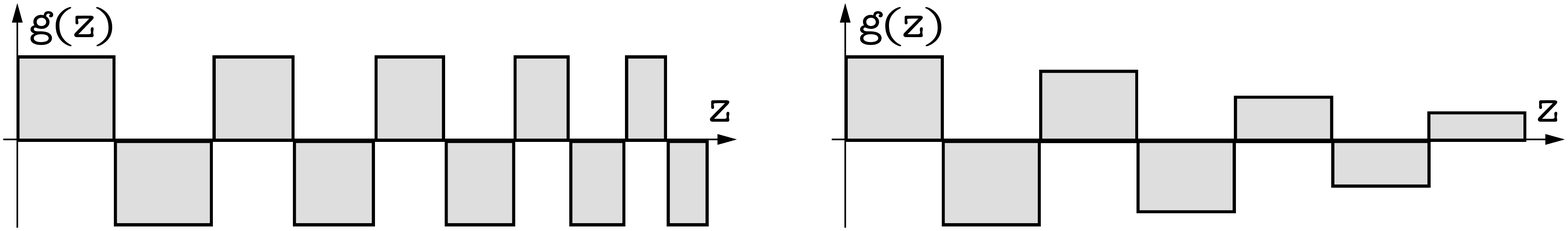}
\includegraphics[scale=0.4]{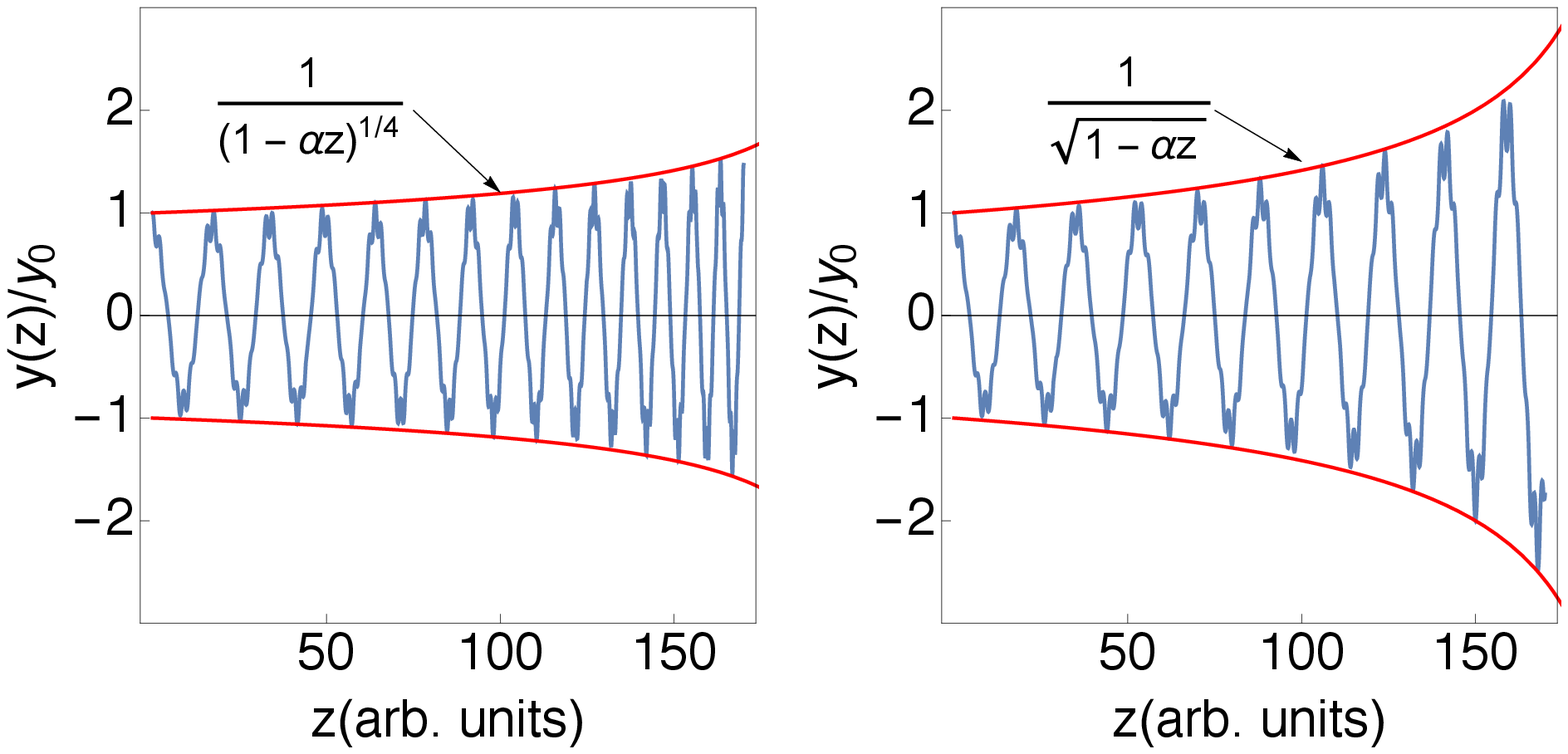}
\caption{Solution of equation \eqref{eq:MainDL} for $s=0$, loss parameter $\alpha=0.005$ and maximum $\sqrt{K}L=0.35\pi$ (phase advance $0.2266\pi$) for adaptive beam focusing using lens length variation $L\propto \sqrt{1-\alpha z}$ (left panel) and lens magnetic gradient variation $g_0\propto(1-\alpha z)$ (right panel).}
\label{Fig:2c}
\end{figure}

By implementing this adaptive focusing and adaptive energy chirp we achieve a condition where the betatron phase advance over each individual FD cell remains the same regardless of the drive bunch energy, and where the beta function decreases in proportion to the length of the cell, in which case the amplitude of betatron oscillation that would normally adiabatically grow as $\mathrm{max}|y|\propto (1-\alpha z)^{-1/2}$ due to the decreasing electron energy, will only grow as $\mathrm{max}|y|\propto (1-\alpha z)^{-1/4}$.
In Fig.\ref{Fig:2c} we show that the quadrupole lens length tapering is better than the tapering of the quadrupole gradient previously considered in \cite{WGai2} and \cite{Li2}.  
Here we plot the solution of equation \eqref{eq:MainDL} for a head particle $s=0$ for the case of adaptive lens length (left panel) and adaptive magnetic gradient (right panel) to illustrate the advantage of the suggested design. 
We would like to emphasize that for adaptive focusing to be valid for the whole bunch, the absolute energy spread $\Delta \gamma$ should be dynamically adjusted so as to keep constant the relative energy spread $\Delta \gamma/ \gamma_z$. 

\section{Two particle model}\label{sec:3}

\begin{figure}[t]
\begin{center}
\includegraphics[scale=0.2]{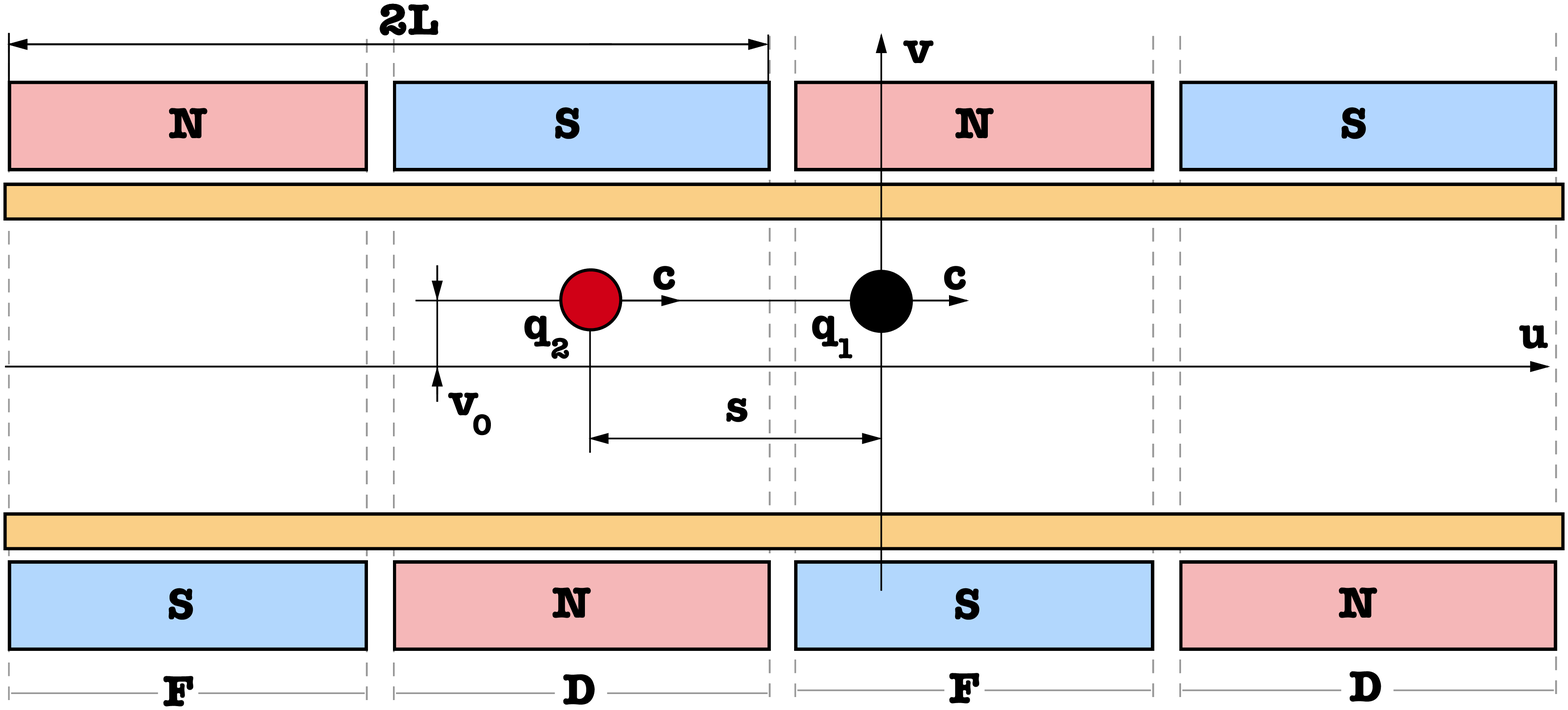}
\caption{A schematic diagram of the two particle model.}
\label{Fig:2d}
\end{center}
\end{figure}

In the subsequent analysis it is more convenient to make substitution in the equation \eqref{eq:hill_r} 
\begin{align}
\label{eq:ncrd}
\frac{1-\tilde{u}}{2\alpha} &\to u, \\
\tilde{v}(s,\tilde{u}) &\to v(s,u), \nonumber
\end{align} 
to obtain
\begin{align}
\label{eq:hill_r2}
&v''(s,u)+\frac{e}{\gamma_0 m_e c} \frac{g(z(u))}{1-f(s)} v(s,u)= \nonumber\\ &=\frac{e}{\gamma_0m_e c^2} \frac{\int\limits_0^s G_1 (s-s_0)q(s_0)v(s_0,u) ds_0}{1-f(s)}.
\end{align}
Now we consider two particles with charges $q_1$ and $q_2$ (see Fig.\ref{Fig:2d}). 
The integro-differential equation \eqref{eq:hill_r2} can be rewritten in this case as a system of two second order differential equations:
\begin{align}
\label{eq:tp1}
v_1''(u)+K(u)v_1(u)=0 ,
\end{align}
\begin{align}
\label{eq:tp2}
v_2''(u)+\frac{K(u)}{1-f(s)}v_2(u)=\frac{w(s)}{1-f(s)}v_1(u).
\end{align}
Here
\begin{align}
\label{eq:KW}
K(u)=\frac{e}{\gamma_0 m_e c}g(z(u)),  \nonumber \\
w(s)=\frac{e}{\gamma_0m_e c^2} G_1 (s)q_1.
\end{align}

Similarly to the BNS damping condition, our goal now is to find such $f(s)$ when the second particle will closely follow the trajectory of the first particle, or in other words, a condition that will keep the distance between second and first particles on a phase space diagram at a minimum defined by the initial conditions:
\begin{align}
\label{eq:BNS}
\argmin\limits_{f(s)}\left[\mathrm{max}\left|\begin{bmatrix} v_2(u) \\ v_2'(u) \end{bmatrix}-\begin{bmatrix} v_1(u) \\ v_1'(u) \end{bmatrix} \right| \right].
\end{align}
In the case of small wakefields, the recipe given by the BNS damping condition \cite{BNS} solves this problem by defining $f(s)$ that produces a small shift of the betatron frequency for the second particle. However, the formalism developed in (\cite{BNS}-\cite{Chao}) lacks applicability in the case of an arbitrary strong wakefield and large shifts of the betatron frequency. We therefore offer a new recipe to define $f(s)$ that continues fulfill condition \eqref{eq:BNS} even in the case of the arbitrary strong wakefields. 

We start with the solution of the equation for the first particle \eqref{eq:tp1} that can be written as
\begin{align}
\label{eq:fpsol}
\begin{bmatrix} v_1(u) \\ v_1'(u) \end{bmatrix}={\mathrm{X}}^u_1\begin{bmatrix} v_{01} \\ v_{01}' \end{bmatrix},
\end{align} 
following \cite{Arnold,Smirnov}. Here $v_{01}$ and $v'_{01}$ are the initial conditions, and ${\mathrm{X}}^u_1$ is the element of the phase flow of equation \eqref{eq:tp1}. Since we consider an FD channel with $K(u)=\pm K$ and $K(u)$ has a period $2L$, we can write the matrix ${\mathrm{X}}^u_1$ in the following form
\begin{align}
\label{eq:pf1}
&{\mathrm{X}}^u_1=\begin{cases} {\mathrm{F}}^u_1 ({\mathrm{A}}_1)^{n-1}, ~~~~~~\mbox{focusing lens,} \\ 
{\mathrm{D}}^u_1{\mathrm{F}}^L_1 ({\mathrm{A}}_1)^{n-1},~~\mbox{defocusing lens}, \end{cases} \\
&~~~~~~~~~~~~~~~u\in\left[0,L\right]\nonumber,
\end{align}
where $n$ is the period number, ${\mathrm{A}}_1={\mathrm{D}}^L_1{\mathrm{F}}^L_1$ is the monodromy matrix or the transfer matrix,
\begin{align}
\label{eq:F1}
{\mathrm{F}}_1^u=\begin{bmatrix} \cos(\sqrt{K}u) && \frac{1}{\sqrt{K}}\sin(\sqrt{K}u) \\ -\sqrt{K}\sin(\sqrt{K}u) && \cos(\sqrt{K}u) \end{bmatrix}
\end{align}

and

\begin{align}
\label{eq:D1}
{\mathrm{D}}_1^u=\begin{bmatrix} \cosh(\sqrt{K}u) && \frac{1}{\sqrt{K}}\sinh(\sqrt{K}u) \\ \sqrt{K}\sinh(\sqrt{K}u) && \cosh(\sqrt{K}u) \end{bmatrix}.
\end{align}
If we assume that initial conditions for the second particle are $v_2(0)=v_{02}$ and $v_2'(0)=v'_{02}$ then the solution of the equation \eqref{eq:tp2} can be found using the free parameters variation method \cite{Arnold} and has the form
\begin{align}
\label{eq:parvar}
&\begin{bmatrix} v_2(u) \\ v_2'(u) \end{bmatrix} = \\ &={\mathrm{X}}_2^u\begin{bmatrix}v_{02} \\ v_{02}' \end{bmatrix}+\frac{w(s)}{1-f(s)}{\mathrm{X}}_2^u\int\limits_{0}^u \left({\mathrm{X}}_2^\tau \right)^{-1} \begin{bmatrix}0 \\ v_1(\tau) \end{bmatrix}d\tau. \nonumber
\end{align}
Here, as before the matrix ${\mathrm{X}}^u_2$, is the element of the phase flow of equation \eqref{eq:tp2} with $w(s)\equiv 0$, that is given by the following equation 
\begin{align}
\label{eq:pf2}
&{\mathrm{X}}^u_2=\begin{cases} {\mathrm{F}}^u_2 ({\mathrm{A}}_2)^{n-1},  ~~~~~~\mbox{focusing lens,}  \\ 
{\mathrm{D}}^u_2{\mathrm{F}}^L_2 ({\mathrm{A}}_2)^{n-1},~~\mbox{defocusing lens}, \end{cases} \\
&~~~~~~~~~~~~~~~u\in\left[0,L\right]\nonumber,
\end{align}
where $n$ is the period number, ${\mathrm{A}}_2={\mathrm{D}}^L_2{\mathrm{F}}^L_2$ is the monodromy matrix or the transfer matrix,
\begin{align}
\label{eq:F2}
{\mathrm{F}}_2^u=\begin{bmatrix} \cos(\theta_2^u) && \sqrt{\frac{1-f(s)}{K}}\sin(\theta_2^u) \\ -\sqrt{\frac{K}{1-f(s)}}\sin(\theta_2^u) && \cos(\theta_2^u) \end{bmatrix}
\end{align}
and
\begin{align}
\label{eq:D2}
{\mathrm{D}}_2^u=\begin{bmatrix} \cosh(\theta_2^u) && \sqrt{\frac{1-f(s)}{K}}\sinh(\theta_2^u) \\ \sqrt{\frac{K}{1-f(s)}}\sinh(\theta_2^u) && \cosh(\theta_2^u) \end{bmatrix}
\end{align} 
with $\theta_2^u=\frac{\sqrt{K}u}{\sqrt{1-f(s)}}$.

We note that we may use in Eq.\eqref{eq:parvar} the same initial conditions for the second particle as for the first particle without losing the generality of the analysis, assuming that there is no wakefield before $u = 0$, in which case one can always find a point on the second particle trajectory before $u = 0$ where $v_2$ and $v'_2$ coordinates are equal to the coordinates of the first particle at $u = 0$. Adding the transfer matrix from that point to $u = 0$ would only make the expressions in the following analysis a bit more cumbersome, but will not add new physics. Therefore, by setting $v_{02}=v_{01}$ and $v'_{02}=v'_{01}$, equation \eqref{eq:parvar} with \eqref{eq:fpsol} can be transformed to
\begin{align}
\label{eq:solsp2}
\begin{bmatrix} v_2(u) \\ v_2'(u) \end{bmatrix} = {\mathrm{T}}^u \begin{bmatrix}v_{01} \\ v_{01}' \end{bmatrix}
\end{align}
with 
\begin{align}
\label{eq:trm}
{\mathrm{T}}^u={\mathrm{X}}_2^u+{\mathrm{X}}_2^u\int\limits_{0}^u \left({\mathrm{X}}_2^\tau \right)^{-1} {\mathrm{W}} {\mathrm{X}}_1^\tau  d\tau
\end{align}
and
\begin{align}
{\mathrm{W}}= \begin{bmatrix}0 && 0 \\ \frac{w(s)}{1-f(s)} && 0 \end{bmatrix}.
\end{align}
 
 We begin by considering the second particle transport through the first focusing lens, i.e., using $n=1$, ${\mathrm{X}}^u_1={\mathrm{F}}^u_1$, ${\mathrm{X}}^u_2={\mathrm{F}}^u_2$, and thus obtaining from Eq. \eqref{eq:trm}: 
\begin{align}
\label{eq:trmf}
{\mathrm{T}}^u={\mathrm{F}}_2^u+{\mathrm{F}}_2^u\int\limits_{0}^u \left({\mathrm{F}}_2^\tau \right)^{-1} {\mathrm{W}} {\mathrm{F}}_1^\tau  d\tau.
\end{align} 
for $u\in[0,L]$.

Performing the matrix multiplication and integration we arrive at the transfer matrix of the second particle after the first focusing lens in the form
\begin{align}
\label{eq:FQ}
{\mathrm{T}}^L=(1-\eta){\mathrm{F}}_2^L+\eta{\mathrm{F}}_1^L,
\end{align} 
here $\eta$ is defined as
\begin{align}
\label{eq:eta}
\eta=\frac{w(s)}{K f(s)}.
\end{align}
Next we achieve a solution in the first defocusing lens from equation \eqref{eq:parvar} with \eqref{eq:FQ} in a form
\begin{align}
	\label{eq:trmf2}
	{\mathrm{T}}^{L+u}={\mathrm{D}}_2^u\left[{\mathrm{T}}^L +\left(\int\limits_{0}^u \left({\mathrm{D}}_2^\tau \right)^{-1} {\mathrm{W}}  {\mathrm{D}}_1^\tau d\tau \right){\mathrm{F}}_1^L\right].
\end{align} 
with $u\in\left[0,L\right]$.

Performing the matrix multiplication and integration we arrive at the transfer matrix for the second particle through the first period in the form 
\begin{align}
\label{eq:tr2}
{\mathrm{T}}^{2L}=(1-\eta){\mathrm{A}}_2-\eta{\mathrm{A}}_1+2\eta{\mathrm{D}}_2^L{\mathrm{F}}_1^L.
\end{align}

One of the possibilities to force the second particle to closely follow the trajectory of the first particle is by requesting 
\begin{align}
\label{eq:trace}
\mathrm{Tr}[{\mathrm{T}}^{2L}]=\mathrm{Tr}[{\mathrm{A}}_1].
\end{align}

By solving equation \eqref{eq:trace} numerically for a given $\sqrt{K}L$ and $\frac{w(s)}{K}$ we may find $f(s)$. The explicit transcendent equation that follows from equation \eqref{eq:trace}  is given in Appendix \ref{sec:app1}.
We also note that the solution of Eq.\eqref{eq:trace} replaces the BNS damping condition for a case of two particles and is not limited to the small wakefields and the small shift of the betatron frequency for the second particle.
Since the motion of both particles is periodic, the above analysis is general and can be repeated with the same conclusion using other periods of the FD channel instead of the first period.

\section{Stability analysis}\label{sec4:intr}

\begin{figure}[h]
\begin{center}
\includegraphics[scale=0.27]{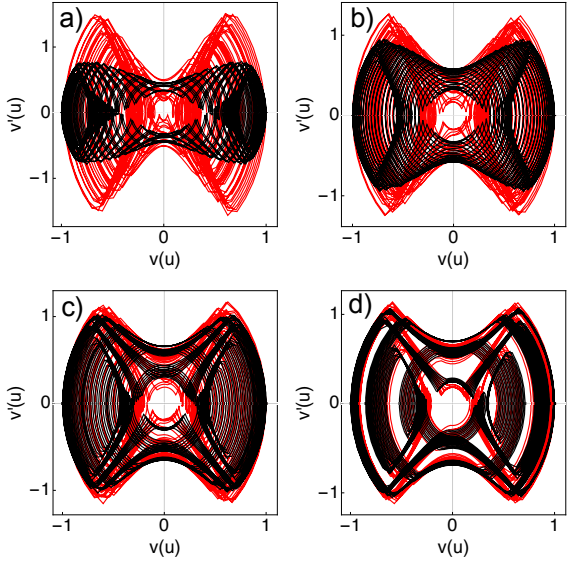}
\caption{Phase trajectories of a first particle (black) and second particle (red). Parameters for the calculation are: Panel a)  $\Phi_1=0.227\pi$, $w(s)/K=0.178$, $f(s)=0.653$; Panel b) $\Phi_1=0.389\pi$, $w(s)/K=0.165$, $f(s)=0.431$; Panel c) $\Phi_1=0.52\pi$, $w(s)/K=0.132$, $f(s)=0.27$; Panel d) $\Phi_1=0.615\pi$, $w(s)/K=0.102$, $f(s)=0.18$. Initial conditions for both particles for all panels are $v_{01}=1$ and $v_{01}'=0$.}
\label{Fig:pp}
\end{center}
\end{figure}

To illustrate the method, we selected four sets of parameters and plotted phase trajectories for the first and second particle in Fig.\ref{Fig:pp}. These parameters were specifically chosen to demonstrate that the method works even at very large energy chirps that may not even be practical. Phase trajectories for the Fig.\ref{Fig:pp} were calculated by a numerical solution of equation \eqref{eq:tp1} and \eqref{eq:tp2} with $f(s)$ determined from Eq.\eqref{eq:trace}.
One can clearly see that phase trajectories are indeed very close to being able to achieve the above formulated goal. Moreover, the maximum coordinate of the second particle is equal to the maximum coordinate of the first particle, i.e.,  $\mathrm{max}|v_2(u)|=\mathrm{max}|v_1(u)|$. 
We also verified that the analytical solutions for phase trajectories given by Eq.\eqref{eq:fpsol}  for the first particle and Eqs.\eqref{eq:solsp2}, \eqref{eq:trm} for the second particle give the same results. In Fig.\ref{Fig:pp} and in subsequent analysis we use the phase advance of the first particle defined as $\Phi_{1}=\mathrm{arccos}\left[\frac{\mathrm{Tr}[{\mathrm{A}}_{1}]}{2}\right]$.

Let us now investigate in more detail the motion of the first and second particles. Stability of the first particle is determined by the eigenvalues of the transfer matrix ${\mathrm{A}}_1$ and stability of the second particle is determined by the eigenvalues of the matrix ${\mathrm{A}}_2$. Thus, the motion of both particles is stable when \cite{Arnold,Smirnov}:
\begin{align}
	\label{eq:pha}
	|{\mathrm{Tr}[{\mathrm{A}}_{1,2}]}|&\leq 2.
\end{align}
We rewrite equation \eqref{eq:eta} with \eqref{eq:KW} as 
\begin{align}
\label{eq:fs1}
f(s)=\frac{1}{\eta(s)}\frac{q_1G_1(s)}{c g_0}
\end{align}    
and analyze the coefficient $1/\eta(s)$.

\begin {figure}[t]
\includegraphics[scale=0.58]{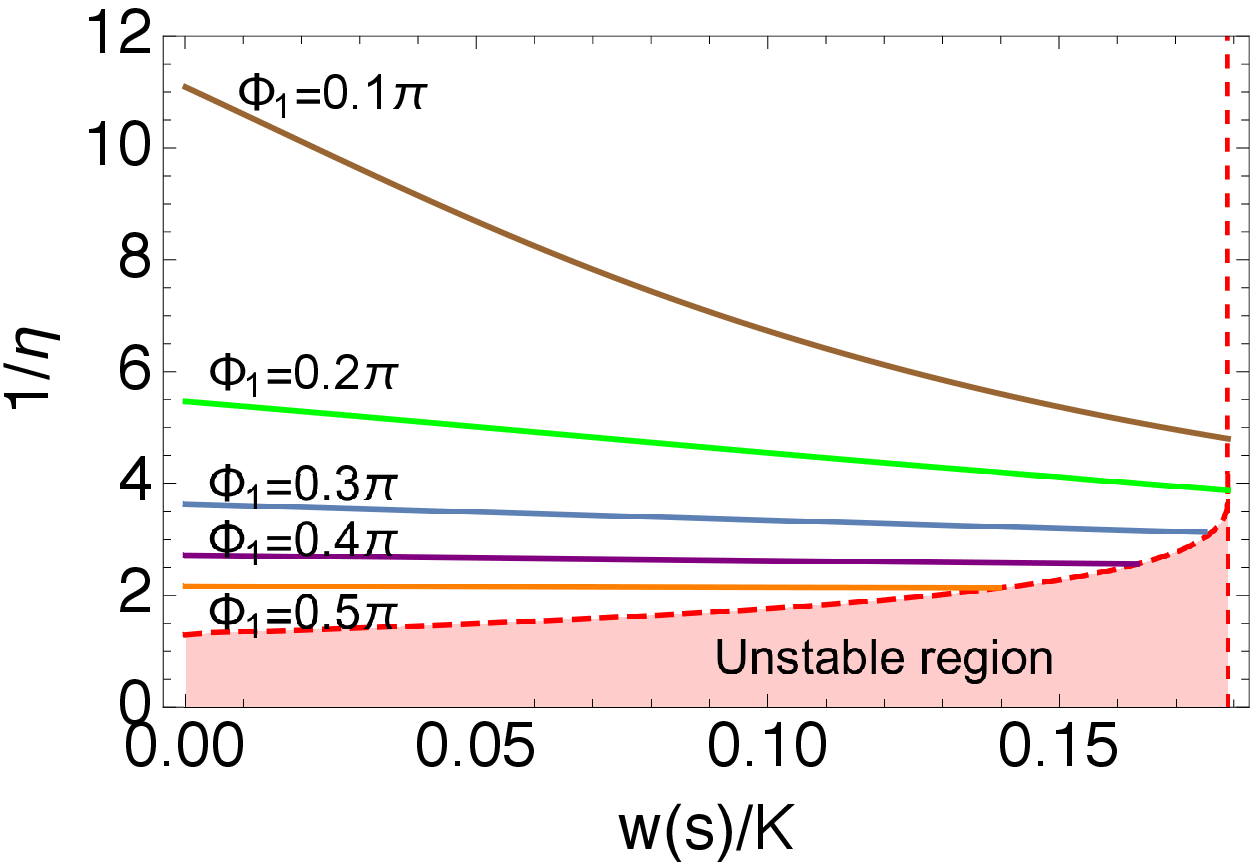}\\
\includegraphics[scale=0.58]{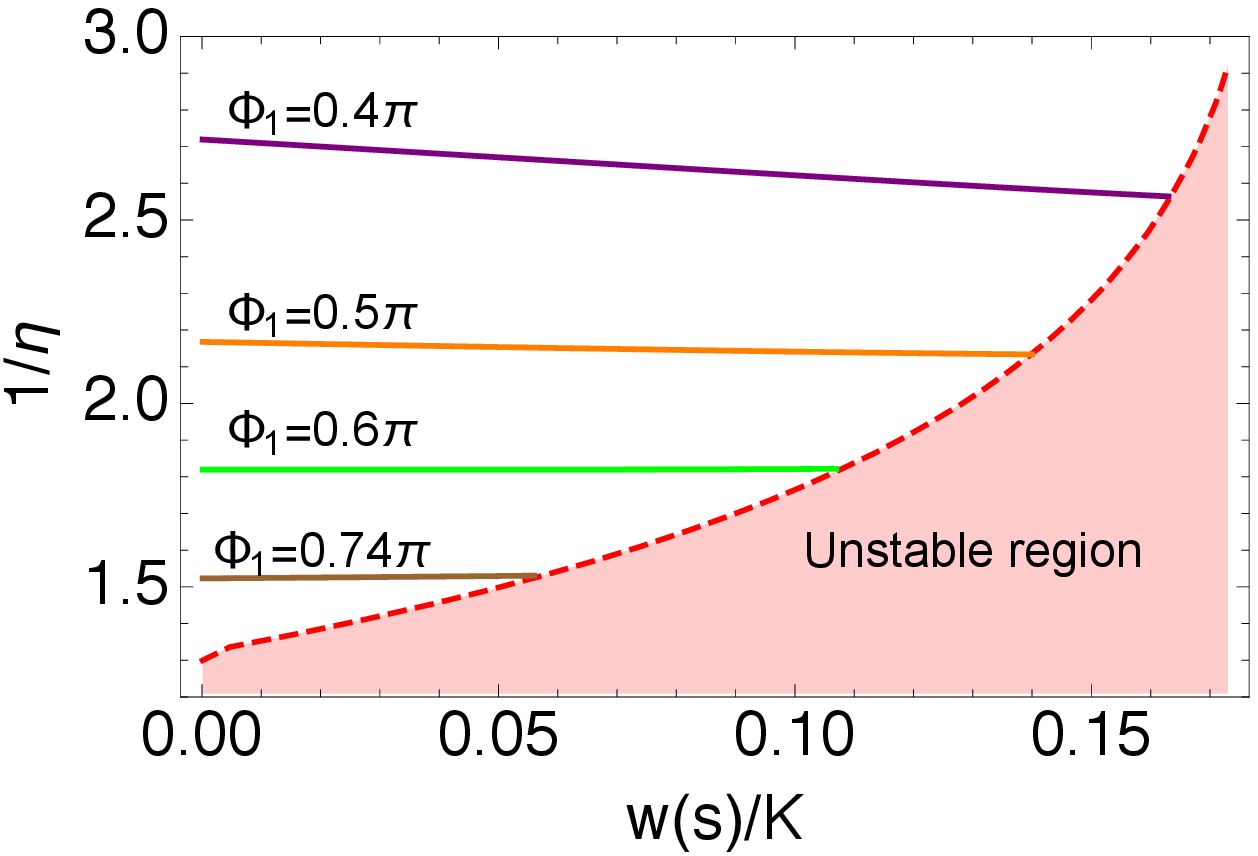}\\
\caption{Coefficient $1/\eta$ calculated using condition \eqref{eq:trace} and formula \eqref{eq:eta} for different values of phase advance $\Phi_1$ of the first particle \eqref{eq:pha} as a function of normalized wakefield; lower panel: zoom of coefficient $1/\eta$ to higher phase advances.} 
\label{fig:etap}
\end {figure}

In Fig.\ref{fig:etap} we plot $1/\eta$ for different values of the phase advance $\Phi_1$ of the first particle
as a function of a normalized wakefield $\frac{w(s)}{K}$ calculated using $f(s)$ found from Eq.\eqref{eq:trace}. One can see that $1/\eta$ greatly increases with the phase advance varying from 0.74$\pi$ to 0.1$\pi$. Consequently, $f(s)$ grows by a large factor, too. Therefore, the lattice with the high value of the phase advance is preferred for a practical CWA to contain the magnitude of the  energy chirp required to stabilize the drive bunch. A shaded area in Fig.\ref{fig:etap} shows the combination of parameters when the motion of the second particle is unstable. This is caused by having an already large frequency of betatron oscillations of the first particle that limits a room to accommodate a frequency shift of the second particle before reaching a condition when  ${|\mathrm{Tr}[{\mathrm{A}}_{2}]}|> 2$. 
As seen in Fig.\ref{fig:etap}, there is some flexibility in selection of the parameters. However, in order to accommodate the maximum ratio of $w(s)/K$ using the minimum energy chirp, we ought to operate in the bottom region of Fig.\ref{fig:etap}. We also note that for phase advance $\Phi_1\geq 0.5 \pi$, the coefficient $1/\eta$ is approximately constant and approximately independent of the wakefield amplitude. Therefore, for a consequent analysis we assume $\Phi_1 \geq 0.5 \pi$ and that coefficient $1/\eta \equiv 1/\eta({\Phi_1})$ is only a function of a phase advance. With this assumption, we rewrite \eqref{eq:fs1} as:
\begin{align}
\label{eq:fs1n}
f(s)=\frac{1}{\eta({\Phi_1})}\frac{q_1G_1(s)}{c g_0}.
\end{align} 
Using formula \eqref{eq:fs1n} and diagram Fig.\ref{fig:etap} one can approximately calculate energy chirp for a given ratio $w(s)/K$ for large phase advances.

It is worth emphasizing that Eq.\eqref{eq:fs1} is very close to a well-know BNS damping condition with the only exception of the coefficient $\eta$ that depends on the wakefield amplitude and can be calculated for a wide range of parameters using Eq.\eqref{eq:trace}.

\section{Stability criteria for the drive bunch in CWA}\label{sec:5}

Now we consider the whole drive bunch and derive the stability condition for its motion in the CWA. 

At first we notice that since $1/{\eta({\Phi_1})}$ is a constant,  we may write for the bunch 
\begin{align}
\label{eq:lim}
f(s)\approx \frac{1}{\eta(\Phi_1)} \frac{\int\limits_0^sG_1(s-s_0)q(s_0)ds_0}{c g_0}.
\end{align}   

Now let us focus on the estimation of the integral in equation \eqref{eq:lim}. At first we write for the decelerating field inside the electron bunch, assuming a single mode longitudinal wakefield
\begin{align}
\label{eq:Ez1}
E_{z}(s,z)=2\kappa_\parallel \int\limits_0^s \cos[k_0(s-s_0)]q(s_0)ds_0,
\end{align}
where $\kappa_{\parallel}$ is the loss factor of a point particle per unit length and $k_0$ is the wave vector of the longitudinal wakefield. By taking the Laplace transformation of $E_{z}(s,z)$ with equation \eqref{eq:Ez}, we obtain for the Laplace image of $q(s)$ 
\begin{align}
\label{eq:Lq}
\tilde{q}(p)=\frac{E_0}{2 \kappa_\parallel} \frac{k_0^2+p^2}{p}\left[\frac{1}{p}-\tilde{f}(p)\right],
\end{align}
where $\tilde{f}(p)$ is the Laplace image of $f(s)$. 

We introduce notation
\begin{align}
\label{eq:int0}
I_1(s)=\int\limits_0^s G_1 (s-s_0)q(s_0) ds_0.
\end{align} 
Next we use a single mode transverse Green's function $G_1(s)=\kappa_\perp/k_1 \sin(k_1s)$, where $\kappa_\perp$ is the kick factor and $k_1$ is the mode's wave vector, and apply the Laplace transformation to obtain:
\begin{align}
\label{eq:int}
\tilde{I}_1(p)=\kappa_{\perp} \frac{\tilde{q}(p)}{p^2+k_1^2}.
\end{align}
Assuming that $\delta=|k_0-k_1|/k_0 <<1$ we rewrite $\tilde{I}_1(p)$ as
\begin{align}
\label{eq:intdec}
\tilde{I}_1(p)=\kappa_{\perp} \tilde{q}(p)\left[\frac{1}{p^2+k_0^2}+\frac{2 k^2_0}{(p^2+k_0^2)^2}\delta+O[\delta]^2 \right],
\end{align}
and using only a zero order term and substitution for $\tilde{q}(p)$ from \eqref{eq:Lq} further obtain: 
\begin{align}
\label{eq:intdec2}
\tilde{I}_1(p)\approx E_0 \frac{\kappa_{\perp}}{2 \kappa_\parallel} \left[\frac{1}{p^2}-\frac{\tilde{f}(p)}{p}\right].
\end{align}
Finally, applying inverse Laplace transformation to \eqref{eq:intdec2} 
we arrive at
 \begin{align}
\label{eq:intdecF}
I_1(s)\approx E_0\frac{\kappa_{\perp}}{2 \kappa_\parallel} \left[s-\int\limits_0^sf(s_0)ds_0\right].
\end{align}
With \eqref{eq:lim} and \eqref{eq:intdecF}  we have
\begin{align}
	\label{eq:Part4}
	f(s)\approx\frac{1}{\eta({\Phi_1})}\frac{E_0}{c g_0 }\frac{\kappa_{\perp}}{2 \kappa_\parallel}\left[s-\int\limits_0^sf(s_0)ds_0\right]. 
\end{align}
Taking the derivative of \eqref{eq:Part4} and introducing $\rho=\frac{E_0}{\eta({\Phi_1}) c g_0 }\frac{\kappa_{\perp}}{2 \kappa_\parallel}$, we find the equation for $f(s)$ 
\begin{align}
	\label{eq:fs}
	f^\prime(s)=-\rho(1-f(s)).
\end{align}
The solution of the equation \eqref{eq:fs} with the initial condition $f(0)=0$ is  
\begin{align}
\label{eq:fsg}
	f(s)=1-e^{\rho s}.
\end{align}
We assume $\rho$ to be is a small parameter. Therefore,
\begin{align}
\label{eq:fsg}
	f(s)=-\rho s-\frac{(\rho s)^2}{2}+O(\rho^3).
\end{align} 
Thus, we have for a linear part of the relative energy variation in the first order to $\rho$:
\begin{align}
	\label{eq:drvBNS}
	\frac{\Delta \gamma}{\gamma_z} \approx -\frac{|E_0|}{\eta(\Phi_1) c g_0}\frac{\kappa_{\perp}}{2 \kappa_\parallel} l.
\end{align}
 Applying the identity $|E_0|=\mathrm{max}|E_+|/R$, where $\mathrm{max}|E_+|$ is the maximum amplitude of the longitudinal electric field behind the electron bunch and $R$ is the transformer ratio, we obtain 
 \begin{align}
	\label{eq:drvBNS1}
	\left|\frac{\Delta \gamma}{\gamma_z}\right| \approx \frac{\mathrm{max}|E_+|}{\eta(\Phi_1) c g_0R}\frac{\kappa_{\perp}}{2 \kappa_\parallel} l.
\end{align}
It was shown in \cite{Baturin} that $R\leq \sqrt{1+k_0^2l^2}$. Consequently, $l\geq \sqrt{R^2-1}/k_0$. With the substitution $g_0=B_0/a_m$, where $B_0$ is the maximum pole tip field for the quadrupole lens with the bore radius $a_m$, we may estimate the amplitude of the energy variation as: 
 \begin{align}
	\label{eq:drvBNS2}
	\left|\frac{\Delta \gamma}{\gamma_z}\right| \gtrsim \frac{a_m}{\eta(\Phi_1)  c B_0 }\frac{\kappa_{\perp}}{ 2\kappa_\parallel}\frac{\mathrm{max}|E_+|}{k_0}\frac{\sqrt{R^2-1}}{R}.
\end{align}
In the cylindrical waveguide with any type of the retarding layer (corrugation, dielectric or semiconductor), the ratio $\kappa_{\perp}/2 \kappa_\parallel$ is strictly bounded by the radius $a_0$ of the waveguide, i.e., $\frac{\kappa_{\perp}}{2 \kappa_\parallel}=\frac{2}{a_0^2}$ (see \cite{myPRL,mySTAB} and references therein). With this substitution and focusing only on the most interesting cases with large transformer ratios, i.e.,  $R>>1$, we reduce Eq.\eqref{eq:drvBNS2} to:

\begin{align}
\label{eq:final}
	\left|\frac{\Delta \gamma}{\gamma_z}\right|  \gtrsim \frac{2}{\eta(\Phi_1)}\frac{a_m}{ k_0a_0^2}\frac{\mathrm{max}|E_+|}{cB_0}.
\end{align} 
The instability region in Fig.\ref{fig:etap} limits the maximum chirp for a given phase advance. We denote the maximum ratio of $w(s)/K$ for a given phase advance $\Phi_1$ as $S({\Phi_1})$ and consequently have  
\begin{align}
\label{eq:final2}
\left|\frac{\Delta \gamma}{\gamma_z}\right|\leq \frac{S({\Phi_1})}{\eta({\Phi_1})}.
\end{align}
Inequalities \eqref{eq:final} and \eqref{eq:final2} 
set boundaries on the energy chirp in the drive bunch required to obtain a certain maximum accelerating field for the witness bunch while maintaining a stable motion of the drive bunch. 

We note that the typical maximum value of $c B_0$ is 300 MV/m defined by the saturation of the magnetic poles in quadrupole lenses. The ratio of the quadrupole bore radius to the radius of the vacuum channel of the wakefield structure is defined by the design constraints of the CWA embedded into the FD channel and here we assume $a_m/a_0=1.5$. To achieve the minimum energy chirp we select the phase advance $\Phi_1=0.74 \pi$ that corresponds to the lowest line for $1/\eta(0.74 \pi)=1.53$ in Fig.\ref{fig:etap}. Consequently, the maximum possible value of $w(s)/K$ is $S(0.74 \pi)=0.056$.   Thus the inequality \eqref{eq:final} can be further rewritten in the engineering form:
\begin{align}
	\label{eq:final_en}
	\left|\frac{\Delta \gamma}{\gamma_z}\right| \gtrsim 1.5\times10^{-2} \frac{\mathrm{max}|E_+| \left(\mathrm{MV/m}\right)}{k_0a_0},
\end{align}
and inequality \eqref{eq:final2}
\begin{align}
\label{eq:final2_en}
\left|\frac{\Delta \gamma}{\gamma_z}\right|\leq 0.086.
\end{align}
Using $\frac{\Delta \gamma}{\gamma_z}$=0.086, $a_0$=1 mm, and $\mathrm{max}|E_+|$=100 MV/m, we find from \eqref{eq:final_en} for a fundamental mode in the CWA $k_0\gtrsim17.4\mathrm{mm}^{-1}$  or 832 GHz. One can see from this example that it is favorable to choose the CWA with the fundamental mode frequency in the THz range to simultaneously achieve a high accelerating gradient for the witness bunch and a stable motion of a drive bunch.

\section{Conclusion}
It has been shown that obtaining a stable motion of the drive bunch in the structure-based collinear wakefield accelerator (CWA) during its deceleration down to a small fraction of its initial energy requires following several important arrangements. The first is the adaptive focusing that gives the benefit of reduction of the adiabatic growth of the amplitude of betatron oscillations from the  $\gamma^{-1/2}$ dependence to $\gamma^{-1/4}$ dependence. The second one is the adaptive energy chirp, i.e., a condition where the longitudinal wakefield is used for continuous reshaping of the initial energy chirp, such as to maintain the same chirp in the relative terms over the entire process of the deceleration of the drive bunch. The last one is actually a prescription describing how to define the initial energy chirp using known parameters of the CWA, a new formalism that is used instead of the BNS damping in the case of strong wakefields. It is based on an extensive analysis of the trajectories of two particles, one driving the wakefield and one subjected to this wakefield, and its extrapolation to the entire drive bunch. It has been shown that when all these arrangements are made to work together, the final result shows that the energy chirp and stability of motion of the drive bunch is defined only by a few parameters combined in one formula. They are the frequency of the wakefield fundamental mode, the radius of the wake inducing retarded medium, the maximum value of the pole tip field of the magnetic lens defined by the saturation of the magnetic poles, and the maximum value of the accelerating field behind the drive bunch. It has also been shown that the criteria for a stable propagation of the drive bunch in the CWA can be satisfied using experimentally realizable parameters. Moreover, these parameters are not tightly constrained and must only fulfill the inequalities \eqref{eq:final} and \eqref{eq:final2}. Finally, a numerical example is given showing a set of realistic parameters allowing achieving a 100 MV/m accelerating gradient for the witness bunch.

\appendix
\section{\label{sec:app1} Transcendent equation for $f(s)$}
Using the fact that trace is a linear operation we may write \eqref{eq:trace}  with \eqref{eq:tr2} and \eqref{eq:eta} as
\begin{align}
\mathrm{Tr}[\mathrm{A}_2]-\mathrm{Tr}[\mathrm{A}_1]+\frac{2w(s)/K}{f(s)-w(s)/K}\mathrm{Tr}[\mathrm{D}_2^L\mathrm{F}_1^L]=0.
\end{align} 
Corresponding traces are found to be
\begin{align}
&\mathrm{Tr}[\mathrm{A}_1]=2\cos\left(\sqrt{K}L\right) \cosh\left(\sqrt{K}L\right), \\
&\mathrm{Tr}[\mathrm{A}_2]=2\cos\left(\frac{\sqrt{K}L}{\sqrt{1-f(s)}}\right) \cosh\left(\frac{\sqrt{K}L}{\sqrt{1-f(s)}}\right),  \nonumber
\end{align}
 and
\begin{align}
&\mathrm{Tr}[\mathrm{D}_2^L\mathrm{F}_1^L]=2 \cos \left(\sqrt{K}L \right) \cosh
   \left(\frac{\sqrt{K}L }{\sqrt{1-f(s)}}\right)+ \nonumber\\&+\frac{f(s) \sin \left(\sqrt{K}L \right) \sinh \left(\frac{\sqrt{K}L
   }{\sqrt{1-f(s)}}\right)}{\sqrt{1-f(s)}}.
\end{align} 

\section{\label{sec:app2} Full numerical simulation of a two particle motion}

To illustrate that all approximations used in the paper are indeed correct, we consider equations of motion of two particles in the most general form
\begin{align}
&\frac{d}{dt}\left[\frac{\gamma_1(t)}{c^2\gamma_0}\frac{dz_1}{dt}\right]=-\alpha_1,  \\
&\frac{d}{dt}\left[\frac{\gamma_1(t)}{c^2\gamma_0}\frac{dy_1}{dt}\right]+K(z_1) y_1=0, \nonumber \\
&\frac{d}{dt}\left[\frac{\gamma_2(t)}{c^2\gamma_0}\frac{dz_2}{dt}\right]=-\alpha_2,  \\
&\frac{d}{dt}\left[\frac{\gamma_2(t)}{c^2\gamma_0}\frac{dy_2}{dt}\right]+\frac{K(z_2)}{1-f(s)}y_2=\frac{w(s)}{1-f(s)}y_1, \nonumber
\end{align}
with relativistic factors $\gamma_{1,2}(t)$ given by  
\begin{align}
\gamma_{1,2}(t)=\frac{1}{\sqrt{1-y_{1,2}'(t)^2/c^2-z_{1,2}'(t)^2/c^2}},
\end{align}
and
\begin{align}
\alpha_{1,2}&=\frac{|e||E_{1,2}|}{\gamma_0m_ec^2}, \nonumber \\
K(z)&=\frac{e}{\gamma_0 m_e c}g(z),  \nonumber \\
w(s)&=\frac{e}{\gamma_0m_e c^2} G_1 (s)q_1.
\end{align}
We solve this equations numerically using parameters listed in Table \ref{tab:1}  and prescriptions given in Sec. \ref{sec:2}. In particular
\begin{align}
\alpha_2=\left[1-f(s)\right]\alpha_1,
\end{align}
\begin {figure}[t]
\includegraphics[scale=0.38]{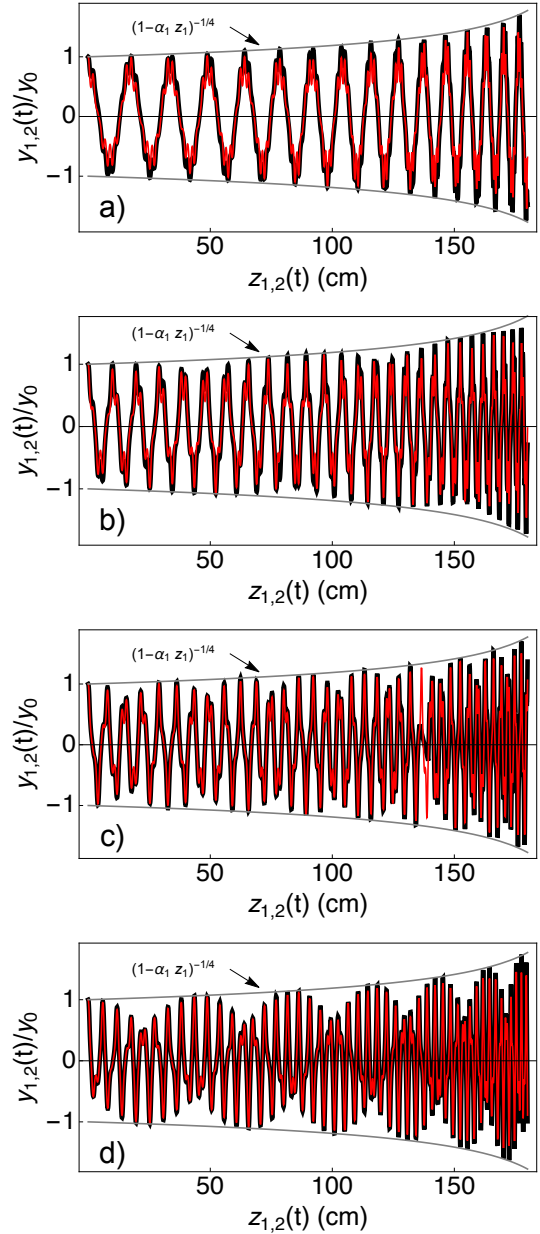}\\
\caption{Trajectories $y_{1,2}(t)$ of a first particle (black) and second particle (red) versus particle longitudinal coordinate $z_{1,2}(t)$ for a) $\Phi_1=0.227\pi$, $w(s)/K=0.12$, $f(s)=0.484$; b) $\Phi_1=0.389\pi$, $w(s)/K=0.13$, $f(s)=0.344$; c) $\Phi_1=0.525\pi$, $w(s)/K=0.08$, $f(s)=0.164$; d) $\Phi_1=0.689\pi$, $w(s)/K=0.056$, $f(s)=0.09$.} 
\label{fig:ap2}
\end {figure}
and adaptive lens length $L(z)=L_0\sqrt{1-\alpha_1 z}$. Simulation was stoped when $\gamma_1(t)=0.1\gamma_0$. To determine $f(s)$ we used condition \eqref{eq:trace}. 

\begin {table}[h!]
\caption{Parameters for the simulation}
\label{tab:1}
\begin{ruledtabular}
\begin {tabular}{c c c c c}
 $\gamma_0$&$\alpha_1~(cm^{-1})$& $y_{1,2}(0)~(cm)$&$y'_{1,2}(0)/c$&$z_{1,2}'(0)/c$ \\
\colrule
$100$&$5\times10^{-3}$ & $3\times10^{-2}$& $0$~&$0.99995$~~ \\
\end {tabular}
\end{ruledtabular}
\end{table}

In Fig.\ref{fig:ap2} we plot trajectories $y_{1,2}(t)$ vs $z_{1,2}(t)$ of the first and second particle for different phase advances $\Phi_1$ of the first particle. We clearly see that for both particles amplitude of the coordinate grows as $\mathrm{max}|y_{1,2}|\propto (1-\alpha_{1} z_{1})^{-1/4}$ and condition \eqref{eq:trace} indeed leads to the equal trajectories $y_{1}(z_1(t))\approx y_{2}(z_2(t))$ during the entire deceleration process.

\section{\label{sec:app2} Approximate analytical formulas for the two particle model}
 
Let us consider equation \eqref{eq:hill_r} for the head particle and write it as 
\begin{align}
\label{eq:hill_p1}
\tilde{v}_1''(\tilde{u})+\frac{4}{\alpha^2} K(\tilde{u}) \tilde{v}_1(\tilde{u})=0.
\end{align}
and for the trailing particle as
\begin{align}
\label{eq:hill_p2}
\tilde{v}_1''(\tilde{u})+\frac{4}{\alpha^2} K(\tilde{u}) \tilde{v}_1(\tilde{u})=\frac{4}{\alpha^2}w(s)\tilde{v}_1(\tilde{u}).
\end{align}
with periodic $K(\tilde{u})$. A
standard approach for an approximate solution is a Floquet transformation and introduction of an average phase advance per FD cell (see for example \cite{Chao}). Since under the conditions of adaptive focusing phase advance $\Phi_1$ is constant, one can write down an approximate solution of \eqref{eq:hill_p1} as
\begin{align}
v_1(\tilde{u})=v_0\cos\left[\frac{\Phi_1}{\alpha L_0} \tilde{u}+\phi_0\right],
\end{align}  
here $\phi_0$ is initial phase and $L_0$ is initial lens length. Applying initial condition $y(0)=y_0$, $y'(0)=0$ and returning back to $z$, variable we have
\begin{align}
\label{eq:s1}
y_1(z)=y_0\frac{\cos\left[\frac{\Phi_1}{\alpha L_0} \sqrt{1-\alpha z}-\frac{\Phi_1}{\alpha L_0}\right]}{\sqrt[4]{1-\alpha z}}.
\end{align}  
Following the same steps and assuming the same initial conditions for the second particle as for the first one, we may write an approximate solution for a wake free ($w(s)\equiv 0$) equation for the second particle \eqref{eq:hill_p2} as
\begin{align}
\label{eq:s2}
y_2(z)=y_0\frac{\cos\left[\frac{\Phi_2}{\alpha L_0} \sqrt{1-\alpha z}-\frac{\Phi_2}{\alpha L_0}\right]}{\sqrt[4]{1-\alpha z}}.
\end{align}  
with $\Phi_{2}=\mathrm{arccos}\left[\frac{\mathrm{Tr}[{\mathrm{A}}_{2}]}{2}\right]$. 
Looking at equation \eqref{eq:tr2} one can come up with an idea of how to construct an approximate solution of \eqref{eq:hill_p2} with the wakefield term in the form
\begin{align}
y_2(z)=\frac{y_0 f_d(s)}{f(s)}\frac{\cos\left[\frac{\Phi_1}{\alpha L_0} \sqrt{1-\alpha z}-\frac{\Phi_1}{\alpha L_0}\right]}{\sqrt[4]{1-\alpha z}}+\\+
y_0\left(1-\frac{f_d(s)}{f(s)} \right)\frac{\cos\left[\frac{\Phi_2}{\alpha L_0} \sqrt{1-\alpha z}-\frac{\Phi_2}{\alpha L_0}\right]}{\sqrt[4]{1-\alpha z}}. \nonumber
\end{align}   
Here $f_d(s)$ is the energy chirp needed for the wakefield cancelation (determined from equation \eqref{eq:trace}) and $f(s)$ is the current energy deviation of a second particle.

\begin{acknowledgments}
This work was supported by the U.S. Department of Energy, Office of Science, Office of Basic Energy Sciences, under Contract No. DE- AC02-06CH11357;  by the U.S. National Science Foundation under Award No. PHY-1549132, the Center for Bright Beams and under Award No. PHY-1535639.   
\end{acknowledgments}

\newpage
\bibliographystyle{ieeetr}

\end{document}